\documentclass[prl,onecolumn,lengthcheck]{revtex4-2}

\usepackage{amssymb, amsmath, amsthm, amstext, amsfonts, amsopn, amscd}
\usepackage{mathrsfs, dsfont, mathdots}
\usepackage{bm}
\usepackage{graphicx}
\usepackage{xcolor}
\usepackage[colorlinks=true, linkcolor=black, breaklinks=true, citecolor=blue, urlcolor=red]{hyperref} 
\usepackage{float}
\usepackage{subcaption}
\usepackage{caption}
\usepackage{tikz}
\usepackage{dcolumn}
\usepackage{todonotes}

\usepackage{pgfplots}
\usetikzlibrary{decorations.markings}
\usetikzlibrary{decorations.pathreplacing,calligraphy}

\definecolor{spacecadet}{HTML}{0D284C}
\definecolor{munsell}{HTML}{008FA8}
\definecolor{banana}{HTML}{FFD932}
\definecolor{cgblue}{HTML}{007CA5}
\definecolor{isabelline}{HTML}{EAEDEA}

\theoremstyle{definition}

\theoremstyle{remark} 
\newtheorem{rem}{Remark}

\newcommand{\Tr}{\operatorname{Tr}}

\newcommand{\cE}{\mathcal E}

\newcommand{\cH}{\mathcal H}

\newcommand{\Z}{\mathbb{Z}}

\newcommand{\1}{\mathds{1}}

\newcommand{\vep}{\varepsilon}

\begin{document}

\title{Anyon braiding and the renormalization group}

\author{Alexander Stottmeister}
\affiliation{$^{1}$Institut f\"ur Theoretische Physik, Leibniz Universit\"at Hannover, Appelstr. 2, 30167 Hannover, Germany}

\begin{abstract}
A braiding operation defines a real-space renormalization group for anyonic chains. The resulting renormalization group flow can be used to define a quantum scaling limit by operator-algebraic renormalization. It is illustrated how this works for the Ising chain, also known as transverse-field Ising model. In this case, the quantum scaling limit results in the vacuum state of the well-known Ising CFT. Distinguishing between the braiding and its inverse is directly related to the chiral sectors of the Ising CFT. This has direct implications for the simulation of CFTs on topological quantum computers.
\end{abstract}

\maketitle


\paragraph{Introduction.} 
Recent tremendous progress in quantum computation \cite{FeynmanLecturesOnComputation, PreskillQuantumComputing40} offers a completely new way to explore complex, strongly-interacting physical systems by quantum simulation, specifically quantum field theories \cite{jordanQuantumAlgorithmsQuantum2012, preskillSimulatingQuantumField2018}. 

A particularly fascinating topic at the interface of quantum field theory and quantum computation is the fractional quantum hall effect \cite{LaughlinAnomalousQuantumHall, StormerTheQuantizedHall} with its possibility of quasiparticles with fractional statistics \cite{BartolomeiFractionalStatisticsIn, NakamuraDirectObservationOf}. In this respect, anyonic chains \cite{FeiguinInteractingAnyonsIn, GilsAnyonicQuantumSpin} have drawn much attention lately. Critical anyonic chains are a potential source of conformal field theories and, thereby, offer specific models of fractional quantum Hall liquids \cite{MooreNonabelionsInThe, ReadBeyondPairedQuantum, ReadPairedStatesOf}.

The relation between critical anyonic chains and their conformal field theories can be understood via scaling limits. The latter arise through the general renormalization group \cite{KadanoffScalingLawsFor, Wilson-71-Renormalization1, Wilson-71-Renormalization2}, one of the key achievements of theoretical physics, offering a unified perspective on complex and subtle phenomena from condensed matter physics \cite{WilsonTheRenormalizationGroupKondo, TetradisCriticalExponentsFrom} over high energy physics \cite{WilsonConfinement, DuerrAbInitioDetermination} to quantum gravity \cite{ReuterEffectiveAverageAction}.

In the context of anyonic chains a notion of scaling limit in the quantum domain is natural \cite{JonesANoGo, ZiniConformalFieldTheories, OsborneContinuumLimitsOf, StottmeisterOperatorAlgebraicRenormalization}. I have recently argued together with Morinelli, Morsella, Tanimoto and Osborne \cite{StottmeisterOperatorAlgebraicRenormalization, MorinelliScalingLimitsOf, OsborneCFTapprox} that a particular approach based on observables and quantum operations, coined operator-algebraic renormalization, provides a practical, conceptually clean definition of quantum scaling limits. This definition directly translates into quantum simulation algorithms, specifically for the limiting quantum field theories \cite{OsborneCFTsim}

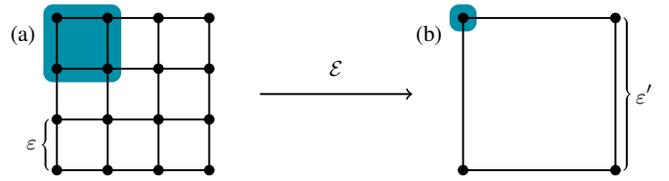
\begin{figure}[h]
\centering
\scalebox{0.9}{
	\begin{tikzpicture}
	
	 
	\draw (-0.5,2) node{(a)};
	
	\def\s{0.2};
	\draw[rounded corners, draw=none, fill=munsell] (0-\s,1.5-\s) rectangle (0.75+\s,2.25+\s) {};
	
	\draw[thick] (0,0) to (2.25,0) to (2.25,2.25) to (0,2.25) to (0,0);
	\draw[thick] (0,0.75) to (2.25,0.75);
	\draw[thick] (0,1.5) to (2.25,1.5);
	\draw[thick] (0.75,0) to (0.75, 2.25);
	\draw[thick] (1.5,0) to (1.5, 2.25);
	\foreach \x in {0,...,3}
	\foreach \y in {0,...,3} 	
	{
	\filldraw (0+0.75*\x,0+0.75*\y) circle (2pt);
	}
	
	\draw[thick, decorate, decoration = {calligraphic brace}] (-0.125,0) --  (-0.125,0.75);
	\draw (-0.375,0.375) node{$\vep$};
	
	
	\draw[thick, ->] (3,1.125) to (5.25, 1.125);
	\draw (4.125,1.5) node{$\cE$};
	
	\draw (5.5,2) node{(b)};
		
	\def\s{0.2};
	\draw[rounded corners, draw=none, fill=munsell] (6-\s,2.25-\s) rectangle (6+\s,2.25+\s) {};
	
	\draw[thick] (6,0) to (8.25,0) to (8.25,2.25) to (6,2.25) to (6,0);
	\foreach \x in {0,1}
	\foreach \y in {0,1} 	
	{
	\filldraw (6+2.25*\x,0+2.25*\y) circle (2pt);
	}
	
	\draw[thick, decorate, decoration = {calligraphic brace}] (8.375,2.25) --  (8.375,0);
	\draw (8.675,1.125) node{$\vep'$};

\end{tikzpicture}
}
\caption{\small Sketch of the block-spin transformation $\cE$ in two dimensions: Neighboring spins placed on a fine lattice (a) with length scale $\vep$ are combined into new averaged block spins on a coarse lattice (b) with length scale $\vep'$.
}
\label{fig:bs}
\end{figure}

The prime example of a real-space renormalization group uses the block-spin transformation as an elementary step \cite{EfratiRealSpaceRenormalization} (see Fig.~\ref{fig:bs}). In this note, I will elaborate on the folklore idea that a braiding operation defines a real-space renormalization group flow on anyonic chains. This can be made precise in the setting of operator-algebraic renormalization and allows for the definition of a quantum scaling limit. In turn this leads to natural quantum simulation algorithms \cite{OsborneCFTsim} on topological quantum computers \cite{KitaevFaultTolerantQuantum, FreedmanTopologicalQuantumComputation}.

The letter is structured as follows: First, I will recall some basic ideas of the renormalization group and its operator-algebraic formulation. After this, I will focus on anyonic chains and describe the basic renormalization group step in terms of a braiding. Then, I will describe a way to construct certain correlation functions between anyons at an arbitrary refinement scale. Following this, I will illustrate the proposed method by constructing the quantum scaling limit of the Ising chain (also known a transverse-field Ising model). Finally, I will offer some conclusions.

\paragraph{Renormalization group and observables.}
The definition of the general renormalization group can be boiled down to setting up a coarse graining operation $\cE$ between states $\rho^{(\vep)}_{0}$ and $\rho^{(\vep')}_{1}$ at fine scale $\vep$ and at coarse scale $\vep'$:
\begin{align}
\label{eq:cg}
\rho^{(\vep')}_{1} & = \cE(\rho^{(\vep)}_{0}) \,.
\end{align}
This defines the renormalization group flow by keeping the coarse scale $\vep'$ fixed and sending the fine scale to zero $\vep\rightarrow 0$. In the terminology of quantum information theory, $\cE$ is a quantum channel, in other words a trace-preserving completely-positive map. By operational duality between states $\rho$ and observables $O$ of a physical system, $\Tr(\cE(\rho)O) = \Tr(\rho\alpha(O))$, we can define a dual refining operation $\alpha$ from the coarse scale $\vep'$ to fine scale $\vep$:
\begin{align}
\label{eq:rf}
O_{\vep} & = \alpha(O_{\vep'}) \,.
\end{align}
The refinement $\alpha$ is a quantum channel on observables, i.e.~a unital completely-positive map. Treating states $\rho$ as expectation functionals $\omega\!=\!\Tr(\rho\!\ \cdot\!\ )$ on observables, the renormalization group flow \eqref{eq:cg} is equivalently written as:
\begin{align}
\label{eq:rgflow}
\omega^{(\vep')}_{1} & = \omega^{(\vep)}_{0}\circ\alpha \,.
\end{align}
As explained in \cite{StottmeisterOperatorAlgebraicRenormalization}, this can be used to define a quantum scaling limit, achieved as follows: One defines a set of limit observables by choosing observables $O_{\vep'}$ at every scale $\vep'$ and iterating \eqref{eq:rf}. In addition, one selects an initial state $\rho^{(\vep)}_{0}$ at every scale $\vep$ and iterates the renormalization group flow \eqref{eq:rgflow}. Correlation functions of limit observables in the quantum scaling limit $\omega_{\infty}$ are computed by:
\begin{align}
\label{eq:qslcor}
\omega_{\infty}(O) & \!=\! \lim_{M\rightarrow\infty}\omega^{(\vep')}_{M}\!\big(O_{\vep'}\big) \!=\! \lim_{M\rightarrow\infty}\!\!\Tr\!\big(\rho^{(\vep')}_{M}O_{\vep'}\big) \, ,
\end{align}
for any choice of scale $\vep'$ and observable $O_{\vep'}$ (converging to $O$ in the sense of \eqref{eq:rf}).

\begin{figure}[h]
		\centering
		\scalebox{1}{
		\begin{tikzpicture}
		\draw[thick] (0,0) to (6,0);
		\draw[thick] (1,0) to (1, 0.5) node[above]{$\sigma$};
		\draw[thick] (2,0) to (2, 0.5) node[above]{$\sigma$};
		\draw[thick] (3,0) to (3, 0.5) node[above]{$\sigma$};
		\draw[thick] (4,0) to (4, 0.5) node[above]{$\sigma$};
		\draw[thick] (5,0) to (5, 0.5) node[above]{$\sigma$};
		\draw (0.5, -0.25) node{$j_{-L}$};
		\draw (1.5, -0.25) node{$\dots$};
		\draw (2.5, -0.25) node{$j_{x}$};
		\draw (3.5, -0.25) node{$\dots$};
		\draw (4.5, -0.25) node{$j_{L-\vep}$};
		\draw (5.5, -0.25) node{$j_{L}$};
		\end{tikzpicture}
		}
	\caption{\small Illustration of an anyonic chain as fusion tree. $\sigma$ labels the anyon and $j_{x}$ are admissible labels in the fusion category.}
	\label{fig:ft}
\end{figure}
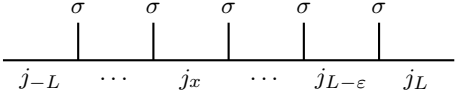

\paragraph{Renormalization group step from braiding.}
Let me explain how the previous framework applies to anyonic chains. The primary input data for an anyonic chain is given by a fusion category \cite{FeiguinInteractingAnyonsIn}. This data is used to define a Hilbert space $\cH_{\vep}$ spanned by basis states $|j_{-L}...j_{x}...j_{L-\vep}\rangle$. The $j_{x}$ are admissible labels of a fusion tree as in Fig.~\ref{fig:ft}.
The positions of the $\sigma$-anyons are labeled by a one-dimensional lattice $\Lambda\!=\!\vep\{\ell,...,\ell\!-\!1\}$. Fixing the total volume of the chain to be finite $L\!=\!\vep\ell$, one defines a Hamiltonian:
\begin{align}
\label{eq:hanyon}
H^{(\vep)} & = \sum_{x\in\Lambda_{N+1}}J_{\vep,x} P^{(0)}_{x} \, .
\end{align}
Each operator $P^{(0)}_{x}$ acts locally on the labels $j_{x-\vep}, j_{x}, j_{x+\vep}$ of the basis states in terms of the $F$-symbols of the fusion category (see \cite[App. C]{GilsAnyonicQuantumSpin} for further details). For anyonic chains based on the fusion rules of $SU(2)_{k}$, it is given by:
\begin{align}
\label{eq:panyon}
& P^{(0)}_{x}|j_{-L}...j_{x}...j_{L-\vep}\rangle \\ \nonumber
& \!=\! \sum_{k_{x}}(F_{j_{x+\vep}}^{j_{x-\vep}\sigma\sigma})^{0}_{j_{x}}(F_{j_{x+\vep}}^{j_{x-\vep}\sigma\sigma})^{0}_{k_{x}}|j_{-L}...k_{x}...j_{L-\vep}\rangle \, .
\end{align}
Here, the Hamiltonian is restricted to a single channel, the ``vacuum'' $0$ (or fusion unit), for simplicity \cite{GilsAnyonicQuantumSpin, WolfMicroscopicModelsOf}. In the $SU(2)_{k}$ case \cite{FeiguinInteractingAnyonsIn}, the operators $P^{(0)}_{x}$ can be identified with generators $e_{x}$ of the Temperley-Lieb algebra $TL(\delta)$\cite{TemperleyRelationsBetweenThe}:
\begin{align}
\label{eq:tl}
e_{x}^{2} & = \delta e_{x}, & e_{x}e_{x\pm\vep}e_{x} & = e_{x}, & e_{x}e_{y} & = e_{y}e_{x} \, ,
\end{align}
where $|x-y|\geq2\vep $ and $\delta$ is the quantum dimension of $\sigma$. 

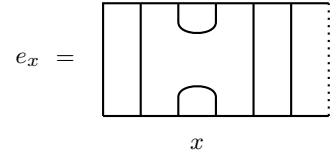
\begin{figure}[h]
		\centering
		\scalebox{1}{
		\begin{tikzpicture}
		
		
		\draw (-1.5,0.75) node{$e_{x}$};
		\draw (-1,0.75) node{$=$};
		\draw[thick] (-0.5,0) to (2.5,0) (2.5,1.5) to (-0.5,1.5) to (-0.5,0);
		\draw[thick] (0,0) to (0,1.5);
		\draw[thick, dotted] (2.5,0) to (2.5,1.5);
		\draw[thick] (0.5,0.25) to[out=90, in=90] (1,0.25) (0.5,0) to (0.5,0.25) (1,0) to (1,0.25);
		\draw[thick] (0.5,1.25) to[out=270, in=270] (1,1.25) (0.5,1.5) to (0.5,1.25) (1,1.5) to (1,1.25);
		\draw[thick] (1.5,0) to (1.5,1.5) (2,0) to (2,1.5);
		\draw (0.75,-0.375) node{$x$};
				
		\end{tikzpicture}
		}
	\caption{\small Diagrammatic representation of a Temperley-Lieb generator $e_{x}$. Strings end at the position of $\sigma$-anyons.}
	\label{fig:tld}
\end{figure}

To develop an intuition, how a real-space renormalization group could act on an anyonic chain, I use the representation of the $e_{x}$ by string box diagrams \cite{KauffmanStateModelsAnd} (see Fig.~\ref{fig:tld}). Strings within a box terminate at the $\sigma$-anyon strings on the chain.
In the following, I assume that observables on the chain are represented by string box diagrams.

\begin{figure}[h]
		\centering
		\scalebox{1}{
		\begin{tikzpicture}
		\draw[thick] (0,0) to (1.5,0) to (1.5,0.5) to (0,0.5) to (0,0);
		\draw[thick] (0.3,0) to (0.3,-0.75) (1.2,0) to (1.2,-0.75); 
		\draw[thick] (0.3,0.5) to (0.3,1.25) (1.2,0.5) to (1.2,1.25); 
		\draw[thick,->] (2,0.25) to (3,0.25);
		\draw (2.5,0.6) node{$\alpha$};
		\draw[thick] (3.5,0) to (5,0) to (5,0.5) to (3.5,0.5) to (3.5,0);
		\draw[thick] (3.8,0) to (3.8,-0.75) (4.7,0) to (4.7,-0.25); 
		\draw[thick] (3.8,0.5) to (3.8,1.25) (4.7,0.5) to (4.7,0.75); 
		\draw[thick, red] (5.6,-0.25) to (5.6,0.75) (6.5,-0.75) to (6.5,1.25); 
		\draw[thick, red] (5.6,-0.25) to[out=270, in=0] (5.4,-0.5) (4.9,-0.5) to[out=180, in=90] (4.7,-0.75);
		\draw[thick] (4.7,-0.25) to[out=270, in=90] (5.6,-0.75);
		\draw[thick, red] (5.6,0.75) to[out=90, in=0] (5.4,1) (4.9,1) to[out=180, in=270] (4.7,1.25);
		\draw[thick] (4.7,0.75) to[out=90, in=270] (5.6,1.25);

		\end{tikzpicture}
		}
	\caption{\small Illustration of the local building block of the renormalization group step $\alpha$. Observables are shown as boxes acting on neighboring anyons, represented as strings. Crossings require a braiding operation.}
	\label{fig:tloar}
\end{figure}
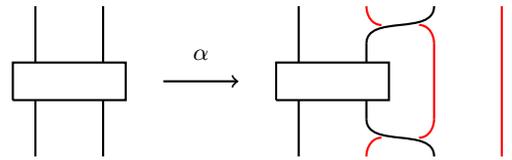

In terms of the diagrammatic representation, the refining renormalization group step $\alpha$ needs to encapsulate the idea that observables on an anyonic chain at a coarse scale $\vep'$ are represented at a fine scale $\vep$ by increasing the local density of strings within in the diagram and on the chain. This corresponds to reading the block-spin transformation in Fig.~\ref{fig:bs} from right to left. Such an operation $\alpha$ can be naturally defined in terms of a braiding because we can simply attach a number of through strings to the left or the right of a given string box diagram. Then, we use the braiding to move the ends of the new string to their required position on the chain, thereby increasing the local density of $\sigma$-anyons. This operation $\alpha$ is depicted for a two-string box diagram in Fig.~\ref{fig:tloar}. The Reidemeister moves \cite{ReidemeisterElementareBegruendungDer} for the diagrammatic calculus \cite{WangTopologicalQuantumComputation} ensure that $\alpha$ is a morphism with respect to the composition of observables given by vertical stacking. Iterating $\alpha$ defines the \textit{braiding renormalization group}.

To compute the renormalization group flow \eqref{eq:rgflow}, we need initial states $\rho^{(\vep)}_{0}$. Sensible candidates are ground states $\rho^{(\vep)}_{0} = |\Omega^{(\vep)}_{0}\rangle\langle\Omega^{(\vep)}_{0}|$ of $H^{(\vep)}$ because the latter is expected to be quantum critical for $J_{\vep,x} = J_{\vep}$. A \textit{quantum scaling limit} is then defined as a limit,
\begin{align}
\label{eq:qsl}
\lim_{\substack{M\rightarrow\infty \\ \vep\rightarrow 0}}\omega^{(\vep)}_{0}\circ\alpha^{M} & = \omega^{(\vep')}_{\infty} \, ,
\end{align}
supplemented by \textit{renormalization conditions} on the coupling constants $J_{\vep,x}$ that ensure an approach to criticality (e.g.~$J_{\vep,x}\rightarrow J_{\vep}$) such that the macroscopic correlation length of observables approaches an intended value (e.g.~finite for massive models, infinite for massless models). The limits $M\rightarrow\infty$, $\vep\rightarrow0$ are taken synchronously, and $\alpha^{M}$ denotes the iterated application of the basic refinement $\alpha$.

\paragraph{Correlation functions.}
A braiding can also be used to define correlation functions and their scaling limit for an anyonic chain. This is achieved in essentially three steps:
\begin{itemize}
	\item[1.] Fix an initial scale $\vep'$ and separate two neighboring, localized anyon observables (or fields) $O_{x}$, $O_{y}$ by successively applying the braiding to one of them.
	\item[2.] Apply the refining channel $\alpha$ to reach the scale $\vep<<\vep'$ while keeping the macroscopic distance of the operators fixed.
	\item[3.] Evaluate the result in a sensible initial state $\rho^{(\vep)}_{0}$ at the refinement scale $\vep$.
\end{itemize}
The procedure is illustrated in Fig.~\ref{fig:tlcor} for the composite operator $P^{(0)}_{x}$ in the $SU(2)_{k}$-Temperley-Lieb setting. $P^{(0)}_{x}$ can be interpreted as the annihilation and subsequent creation of a pair of neighboring anyons at the coarse scale.

\begin{widetext}

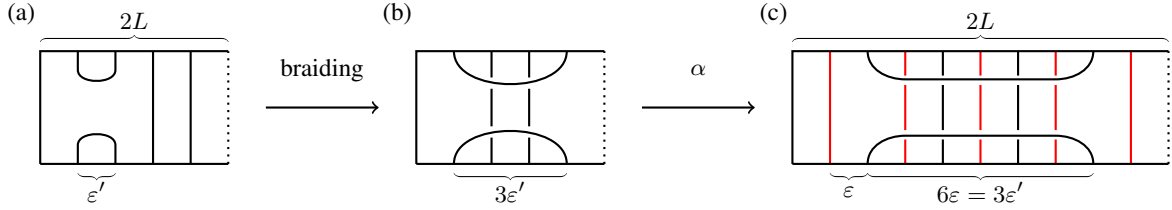
\begin{figure}[h]
		\centering
		\scalebox{1}{
		\begin{tikzpicture}
		
		
		\draw (-0.25,2) node{(a)};
		
		\draw[decorate, decoration = {calligraphic brace}] (0,1.625) --  (2.5,1.625);
		\draw (1.25,1.9) node{$2L$};
		\draw[thick] (0,0) to (2.5,0) (2.5,1.5) to (0,1.5) to (0,0);
		\draw[thick, dotted] (2.5,0) to (2.5,1.5);
		\draw[thick] (0.5,0.25) to[out=90, in=90] (1,0.25) (0.5,0) to (0.5,0.25) (1,0) to (1,0.25);
		\draw[thick] (0.5,1.25) to[out=270, in=270] (1,1.25) (0.5,1.5) to (0.5,1.25) (1,1.5) to (1,1.25);
		\draw[thick] (1.5,0) to (1.5,1.5) (2,0) to (2,1.5);
		\draw[decorate, decoration = {calligraphic brace}] (1,-0.125) --  (0.5,-0.125);
		\draw (0.75,-0.375) node{$\vep'$};
		
		
		\draw[thick, ->] (3,0.75) to (4.5,0.75);
		\draw (3.75,1.25) node{braiding};
		
		
		\draw (4.75,2) node{(b)};
		
		\draw[thick] (6,0) to (6,0.35) (6,0.5) to (6,1) (6,1.15) to (6,1.5);
		\draw[thick] (6.5,0) to (6.5,0.35) (6.5,0.5) to (6.5,1) (6.5,1.15) to (6.5,1.5);
		\draw[thick] (5,0) to (7.5,0) (7.5,1.5) to (5,1.5) to (5,0);
		\draw[thick, dotted] (7.5,0) to (7.5,1.5);
		\draw[thick] (5.5,0) to[out=90, in=90] (7,0);
		\draw[thick] (5.5,1.5) to[out=270, in=270] (7,1.5);
		\draw[decorate, decoration = {calligraphic brace}] (7,-0.125) --  (5.5,-0.125);
		\draw (6.25,-0.375) node{$3\vep'$};
		
		
		\draw[thick, ->] (8,0.75) to (9.5,0.75);
		\draw (8.75,1.25) node{$\alpha$};
		
		
		\draw (9.75,2) node{(c)};
		
		\draw[decorate, decoration = {calligraphic brace}] (10,1.625) --  (15,1.625);
		\draw (12.5,1.9) node{$2L$};
		\draw[thick, red] (10.5,0) to (10.5,1.5);
		\draw[thick, red] (11.5,0) to (11.5,0.325) (11.5,0.45) to (11.5,1.05) (11.5,1.175) to (11.5,1.5);
		\draw[thick] (12,0) to (12,0.325) (12,0.45) to (12,1.05) (12,1.175) to (12,1.5);
		\draw[thick, red] (12.5,0) to (12.5,0.325) (12.5,0.45) to (12.5,1.05) (12.5,1.175) to (12.5,1.5);
		\draw[thick] (13,0) to (13,0.325) (13,0.45) to (13,1.05) (13,1.175) to (13,1.5);
		\draw[thick, red] (13.5,0) to (13.5,0.325) (13.5,0.45) to (13.5,1.05) (13.5,1.175) to (13.5,1.5);
		\draw[thick, red] (14.5,0) to (14.5,1.5);
		\draw[thick] (10,0) to (15,0) (15,1.5) to (10,1.5) to (10,0);
		\draw[thick, dotted] (15,0) to (15,1.5);
		\draw[thick] (11,0) to[out=90, in=180] (11.5,0.375) (11.5,0.375) to (13.5,0.375) (14,0) to[out=90, in=0] (13.5,0.375);
		\draw[thick] (11,1.5) to[out=270, in=180] (11.5,1.125) (11.5,1.125) to (13.5,1.125) (14,1.5) to[out=270, in=0] (13.5,1.125);
		\draw[decorate, decoration = {calligraphic brace}] (14,-0.125) --  (11,-0.125);
		\draw (12.5,-0.375) node{$6\vep=3\vep'$};
		\draw[decorate, decoration = {calligraphic brace}] (11,-0.125) --  (10.5,-0.125);
		\draw (10.75,-0.375) node{$\vep$};
		
		\end{tikzpicture}
		}
	\caption{\small Construction of a correlation function at scale $\vep$ on an anyonic chain of physical length $2L$. The boxes represent a local operator $P^{(0)}_{x}$ modeling the annihilation and subsequent creation of two anyons. Initially (a), the anyons are located at neighboring sites on the coarse chain, separated by a distance $\vep'$. Intermediately (b), the braiding is used to separate the anyons at a macroscopic distance $3\vep'$. Finally (c), the chain is refined by the quantum channel $\alpha$ keeping the anyons at a fixed macroscopic distance $6\vep = 3\vep'$.}
	\label{fig:tlcor}
\end{figure}

\end{widetext}

\begin{rem}
If the sites of an anyonic chain are indexed by dyadic rationals, the methods presented in \cite{BrothierConstructionsOfConformal} can be used to define a Jones representation of Thompson's group $F$ (and $T$ for periodic chains) \cite{JonesSomeUnitaryRepresentations, JonesANoGo} on the limit observables. It is tempting to assume that elements of $F$ are discrete approximations of conformal transformations. While such an approximation may hold at the level of observables, it is obstructed at the level of Hilbert space (i.e.~for correlation functions) by smoothness constraints \cite{DelVecchioSolitonsAndNonsmooth}.
\end{rem}

\paragraph{The Ising chain.}
Let me briefly explain how the quantum scaling limit in the vacuum sector of the critical Ising chain can be explicitly computed using the braiding refinement $\alpha$.

A sensible basic algebra of observables for the Ising chain at (log-)scale $N\!=\!-\log_{2}(\vep_{N}/\vep_{0})$ is the periodic Temperley-Lieb algebra $TL_{N+1}(\delta)$ with $2^{N+1}$ generators and loop parameter $\delta\!=\!\sqrt{2}$ \cite{GrimmTheSpin12, FeiguinInteractingAnyonsIn}. Here, the generators labeled by the dyadic lattice $\Lambda_{N+1}$, $\ell\!=\!L_{N+1}=2^{N+1}L_{0}$ or concreteness. It is well-known that $TL_{N+1}(\delta)$ can realized in terms of a Majorana fermion $\{\psi_{x},\psi_{y}\}\!=\!2\delta_{x,y}$  with anti-periodic boundary conditions $\psi_{L}\!=\!-\psi_{-L}$ \cite{SchuetzDualityTwistedBoundary}:
\begin{align}
\label{eq:tlmaj}
e_{x} &\!=\!\tfrac{1}{\sqrt{2}}(1\!+\!i\psi_{x+\vep_{N+1}}\psi_{x}) \, .
\end{align}
The (unitary) braiding is provided by the Kauffman bracket \cite{KauffmanStateModelsAnd},
\begin{figure}[h]
		\centering
		\scalebox{1}{
		\begin{tikzpicture}
		\draw (-0.75,0.25) node{$b_{x}$};
		\draw (-0.375,0.25) node{$=$};
		\draw[thick] (0,0) to[out=90, in=270] (0.5,0.5);
		\draw[thick] (0.5,0) to[out=90, in=330] (0.275,0.225) (0.225,0.275) to[out=150, in=270] (0,0.5);
		\draw (0.875,0.25) node{$=$};
		\draw (1.25,0.25) node{$A$};
		\draw[thick] (1.5,0) to (1.5,0.5);
		\draw[thick] (2,0) to (2,0.5);
		\draw (2.3,0.3) node{$+$};
		\draw (2.85,0.3) node{$A^{-1}$};
		\draw[thick] (3.25,0.5) to[out=270, in=270] (3.75,0.5);
		\draw[thick] (3.25,0) to[out=90, in=90] (3.75,0);
		\draw (4.025,0.25) node{$=$};
		\draw (4.25,0.3) node[right]{$A+A^{-1}e_{x}$};
		\end{tikzpicture}
		}
	\caption{\small The Kauffman bracket \cite{KauffmanStateModelsAnd} representing the braiding in terms of Temperley-Lieb generators. The parameter $A$ is related to the loop parameter $\delta$ by: $\delta = -A^{2}-A^{-2}$.}
	\label{fig:tlbk}
\end{figure}

or (up to a phase) by the Jones representation \cite{JonesAPolynomialInvariant}:
\begin{align}
\label{eq:tlbj}
b_{x} &\!=\!\tfrac{i}{\sqrt{2}}(\psi_{x+\vep_{N+1}}\psi_{x}\!-\!1) \, .
\end{align}
As depicted in Fig.~\ref{fig:tlcor}, (b), the braiding can be used to separate two neighboring fermions (understood as fields on the Ising chain):
\begin{align}
\label{eq:tlbact}
b_{x}\psi_{x}b_{x}^{*} &\!=\! -\psi_{x+\vep_{N+1}}, & b_{x}\psi_{x+\vep_{N+1}}b_{x}^{*} &\!=\! \psi_{x}, \, .
\end{align}
By successive conjugation of the Temperley-Lieb generators $e_{x}$ with $b_{x-\vep_{N+1}}$, $b_{x-2\vep_{N+1}}$ and so forth, we can produce any fermion bilinear $\psi_{x}\psi_{y}$ with $|x-y|=n\vep_{N}$.

The Hamiltonian of the critical anti-periodic Ising chain at scale $N$ is given in terms of $TL_{N+1}(\delta)$ \cite{SchuetzDualityTwistedBoundary} as:
\begin{align}
\label{eq:hising}
H^{(N)}_{0} & = J\sum_{x\in\Lambda_{N+1}}e_{x} \, .
\end{align}
The many-body ground state $\Omega^{(N)}_{0}$ of $H^{(N)}_{0}$ is fully determined by two-fermion expectation values (see e.g.~\cite{OsborneCFTapprox}):
\begin{align}
\label{eq:hgs}
\omega^{(N)}_{0}\!(\psi_{x}\psi_{y}) & \!=\! \langle\Omega^{(N)}_{0}|\psi_{x}\psi_{y}|\Omega^{(N)}_{0}\rangle \\ \nonumber
& \!=\! \tfrac{i}{2L_{N}}\!\!\sum_{k\in\Gamma_{N,+}}\!\!\!\sin(k(x\!-\!y)\!)\tfrac{\sin(\vep_{N}k)}{\sin(\frac{1}{2}\vep_{N}k)} \, .
\end{align}
A direct computation shows that the $M$-fold iteration of \eqref{eq:cg} at scale $N$ using the scheme in Fig.~\ref{fig:tlcor} yields:
\begin{align}
\label{eq:hgsren}
\omega^{(N)}_{M}\!(\psi_{x}\psi_{y}) &\!=\! \tfrac{i}{2L_{N}}\!\!\!\!\!\sum_{k\in\Gamma_{N+M,+}}\!\!\!\!\!\!\sin(k(x\!-\!y)\!)\tfrac{\sin(\vep_{N+M}k)}{\sin(\frac{1}{2}\vep_{N+M}k)} \, .
\end{align}
For simplicity, I included a factor $2^{M}$ in the refining channel $\alpha$ from scale $N$ to $N+M$ to account for the canonical scaling dimension of $\psi_{x}\psi_{y}$. From \eqref{eq:hgsren}, it is easy to read of limit the renormalization group flow:
\begin{align}
\label{eq:hgsrenlim}
\omega^{(N)}_{\infty}\!(\psi_{x}\psi_{y}) &\!=\! \tfrac{i}{L_{N}}\!\!\!\sum_{k\in\frac{\pi}{L}\Z_{\geq0}}\!\!\!\!\sin(\!(k\!+\!\tfrac{\pi}{2L})(x\!-\!y)\!) \, .
\end{align}
This is the expected correlation function of the $c=\tfrac{1}{2}$ Ising CFT on a circle of circumference $2L$ \cite{DiFrancescoCFTBook} which also gives the correct result in the infinite-volume vacuum state $\omega^{(N)}_{\infty}\rightarrow\omega^{(N)}$, $L\rightarrow\infty$, at scale $N$:
\begin{align}
\label{eq:hgsrenliminf}
\omega^{(N)}\!(\psi_{x}\psi_{y}) &\!=\! \tfrac{i\vep_{N}}{\pi}\Big(\tfrac{1}{x-y+i0}+\tfrac{1}{x-y-i0}\Big) \, .
\end{align}
There is a residual factor $\vep_{N}$ due to the scaling dimension of $\psi$ and the fact that the limit was taken for fixed (but arbitrary) $N$. The two-point function \eqref{eq:hgsrenliminf} explicitly shows that the quantum scaling limit correctly restores conformal invariance. By the reasoning below \eqref{eq:tlbact}, it is also possible to express \eqref{eq:hgsrenlim} in terms of the Temperley-Lieb generators at an arbitrary scale $N$:
\begin{align}
\label{eq:hgsrenliminftl}
\omega^{(N)}\!(\sqrt{2}e_{x}-1) &\!=\! \tfrac{i\vep_{N}}{\pi}\Big(\tfrac{1}{\vep_{N+1}+i0}+\tfrac{1}{\vep_{N+1}-i0}\Big) \, ,
\end{align}
which is independent of $x$ by translation invariance. The length $\vep_{N+1}$ is interpreted as macroscopic length in the scaling limit.

\paragraph{Chirality.}
Note that the standard chiral and anti-chiral components $\psi_{\pm}$ of the Ising model \cite{SatoHolonomicQuantumFieldsIV, SatoHolonomicQuantumFieldsV} are related to the Majorana fermion by:
\begin{align}
\label{eq:chiral}
\psi_{\pm|x} & = \tfrac{1}{\sqrt{2}}(\psi_{x}\mp\psi_{x+\vep_{N+1}}) \, .
\end{align}
This means that the Majorana components associated with even and odd sites of the chain at finite scale equivalently capture the two chiral sectors. It is evident from the pictorial representation of the basic refinement $\alpha$ in Fig.~\ref{fig:tloar} that the quantum scaling limit extracts the component of continuum Majorana fermion corresponding to either even or odd sites at finite scale. More precisely: The continuum component of the even sites will appear if the ends of new trough strings (red) are attached to the right of the original anyon sites. The continuum component of the odd sites will appear if the news ends are attached to the left of the original sites.
Thus, building $\alpha$ from either the braiding or its inverse according to \eqref{eq:tlbact}, i.e.~separating neighboring fermions by moving to the left or right, is directly related to which (linear combination of) chiral components of the Majorana fermion are found in the quantum scaling limit.

\paragraph{Conclusion.}
I have used the idea that a braiding leads to a natural renormalization group flow on anyonic chains and made this precise using operator-algebraic renormalization. This flow defines a quantum scaling limit, given an initial state $\rho^{(N)}_{0}$, exhibiting criticality, at every scale $N$. The braiding renormalization group can be explicitly realized for $SU(2)_{k}$ quantum spin chains or related ones, such as the Golden Chain \cite{FeiguinInteractingAnyonsIn}, by exploiting their Temperley-Lieb algebra structure. The definition of the refining channel $\alpha$ applies to open chains as well.

The braiding of anyons is a natural operation in the topological approach to quantum computation \cite{FreedmanTopologicalQuantumComputation, WangTopologicalQuantumComputation}. The concept of quantum scaling limit presented here leads, in combination with \cite{OsborneCFTapprox}, to simulation algorithms for quantum field theories on topological quantum computers.

The correlation functions sketched in Fig.~\ref{fig:tlcor} can be used as a new tool, besides the entanglement entropy \cite{HolzheyGeometricAndRenormalized, CalabreseEntanglementEntropyAnd}, to numerically investigate anyonic chains and possibly aide the search for new conformal field theories, including exotic ones \cite{HuangNumericalEvidenceFor, VanhoveACriticalLattice}.

As an example, I have shown how this renormalization group can be used to define a quantum scaling limit of the Ising chain. The resulting vacuum state is directly related to the chiral halves of the $c=\tfrac{1}{2}$ Ising CFT. This is another direct proof of the circumvention of the no-go result of Jones \cite{JonesANoGo} using operator-algebraic renormalization, as first observed in \cite{StottmeisterOperatorAlgebraicRenormalization}. A similar observation was made by Zini and Wang in \cite{ZiniConformalFieldTheories}, using the notion of a low-energy scaling limit. A more detailed account of the relation between their approach and the method presented here can be found in \cite{OsborneCFTapprox}.
\begin{acknowledgments}
I thank T.~J.~Osborne, A.~Hahn, R.~Wolf and T.~Cope for interesting and helpful discussions. I also thank A.~Brothier for discussions about the braiding at an early stage of the research for this paper during the workshop ``Operator Algebras and Applications'', June 17-21, 2019 at the Simons Center for Geometry and Physics. Concerning the latter, I gratefully acknowledge support from the Simons Center for Geometry and Physics, Stony Brook University.
\end{acknowledgments}


%

\end{document}